\documentclass[prb,aps,twocolumn,superscriptaddress,nofootinbib]{revtex4-1}
\pdfoutput=1
\usepackage{amssymb,graphicx,color}
\usepackage[intlimits]{amsmath}
\usepackage[english]{babel}
\usepackage[colorlinks]{hyperref}
\usepackage{comment}
\usepackage{tabularx}
\usepackage{array}
\usepackage[svgnames,table]{xcolor}
\usepackage{ragged2e}

\newcolumntype{A}{>{\centering\arraybackslash \columncolor{white!50!white}}m{2.1cm}}
\newcolumntype{B}{>{\centering\arraybackslash \columncolor{white}}m{7.9cm}}
\newcolumntype{C}{>{\centering\arraybackslash \columncolor{white!50}}m{7.9cm}}
\newcolumntype{D}{>{\centering\arraybackslash \columncolor{white!42}}m}
\newcolumntype{P}[1]{>{\centering\arraybackslash}p{#1}}
\def\beq{\begin{equation}}
\def\eeq{\end{equation}}
\def\bea{\begin{eqnarray}}
\def\eea{\end{eqnarray}}
\def\barr{\begin{array}}
\def\earr{\end{array}}

\begin{document}

\title{Entanglement and thermodynamic entropy in a clean many-body localized system}

\author{Devendra Singh Bhakuni}
\affiliation{Indian Institute of Science Education and Research Bhopal 462066 India}
\author{Auditya Sharma}
\affiliation{Indian Institute of Science Education and Research Bhopal 462066 India}

\begin{abstract}
Whether or not the thermodynamic entropy is equal to the entanglement
entropy of an eigenstate, is of fundamental interest, and is closely
related to the `Eigenstate thermalization hypothesis (ETH)'. However,
this has never been exploited as a diagnostic tool in many-body
localized systems. In this work, we perform this diagnostic test on a clean interacting system
(subjected to a static electric field) that exhibits three distinct
phases: integrable, non-integrable ergodic and non-integrable
many-body-localized (MBL).  We find that in the non-integrable phase,
the equivalence between the thermodynamic entropy and the entanglement
entropy of individual eigenstates, holds. In sharp contrast, in the
integrable and non-integrable MBL phases, the entanglement entropy
shows large eigenstate-to-eigenstate fluctuations, and differs from the
thermodynamic entropy. Thus the non-integrable
MBL phase violates ETH similar to an integrable system; however, a key
difference is that the \emph{magnitude} of the entanglement entropy in
the MBL phase is significantly smaller than in the
integrable phase, where the entanglement entropy is of the same order
of magnitude as in the non-integrable phase, but with a lot of
eigenstate-to-eigenstate fluctuations. Quench dynamics from an initial CDW state
independently supports the validity of the ETH in the ergodic phase
and its violation in the MBL phase.
\end{abstract}

\maketitle
\section{Introduction} The question of how an isolated
many-body system thermalizes has a long history. In the classical
domain, thermalization of an isolated system in the limit of long
times is governed by Boltzmann's ergodic
hypothesis~\citep{huang2009introduction,pathria2011statistical,penrose1979foundations}. It
states that classical chaotic systems, uniformly sample all the
available micro-states at a given energy, in the long time
limit. However, this hypothesis cannot be generalized directly to the
quantum domain as in the long time limit the expectation value of an
observable retains the initial memory of the system, and is thus
unable to sample all the eigenstates of the system. Experimental
advancement~\cite{kinoshita2006quantum,hofferberth2007non,rigol2008thermalization}
in recent times has created a strong demand for a close understanding
of thermalization in isolated quantum systems and led to a flurry of
theoretical activity~\citep{PhysRevLett.103.100403,PhysRevE.82.031130,rigol2011initial,he2012initial,mondaini2015many,PhysRevE.89.042112,rigol2007relaxation,PhysRevE.81.036206,canovi2011quantum,rigol2010quantum}.

Thermodynamic entropy in the context of classical statistical
mechanics is by its very nature an extensive
quantity~\citep{huang2009introduction,penrose1979foundations,pathria2011statistical}. In
quantum systems, entanglement entropy of individual eigenstates brings
in a rich additional dimension. Discussions of the extensivity or the
lack thereof of entanglement entropy have
abounded~\citep{vitagliano2010volume,PhysRevLett.96.010404,PhysRevLett.111.210402,roy2018entanglement,roy2019quantum,PhysRevB.89.115104}
in recent times. The celebrated area
law~\citep{hastings2007area,laflorencie2016quantum,eisert2010colloquium}
which asserts that the ground state entanglement entropy scales with
subsystem as the surface area of the subsystem, has been a central
topic around which many of these studies have been carried
out. However, the relationship between entanglement entropy and
thermodynamic entropy has only been scantily
covered~\cite{deutsch2013microscopic}. In this Letter, we demonstrate,
with the aid of a specific example, that a systematic study of this
relationship is an illuminating diagnostic for a class of quantum
phase transitions.

For an isolated quantum system it has been argued that the route to
thermalization is described by the \emph{eigenstate thermalization
  hypothesis}(ETH)\citep{PhysRevA.43.2046,PhysRevE.50.888,rigol2008thermalization,deutsch2018eigenstate,d2016quantum}. The
ETH states that expectation values of operators in the eigenstates of
the Hamiltonian are identical to the their thermal values, in the
thermodynamic limit.  The measurement of any local observable in these
systems gives the same expectation values for nearby energies. A
closely related, but completely independent feature analogous to the ETH
is the question of whether the thermodynamic entropy of a subsystem
obtained from the micro-canonical reduced density matrix with a fixed
energy $E_0$ is equal to the entanglement entropy calculated from the
energy eigenstate of the system with the same energy
$E_0$~\citep{deutsch2010thermodynamic,deutsch2013microscopic}.

\begin{figure}[t]
\includegraphics[scale=0.65]{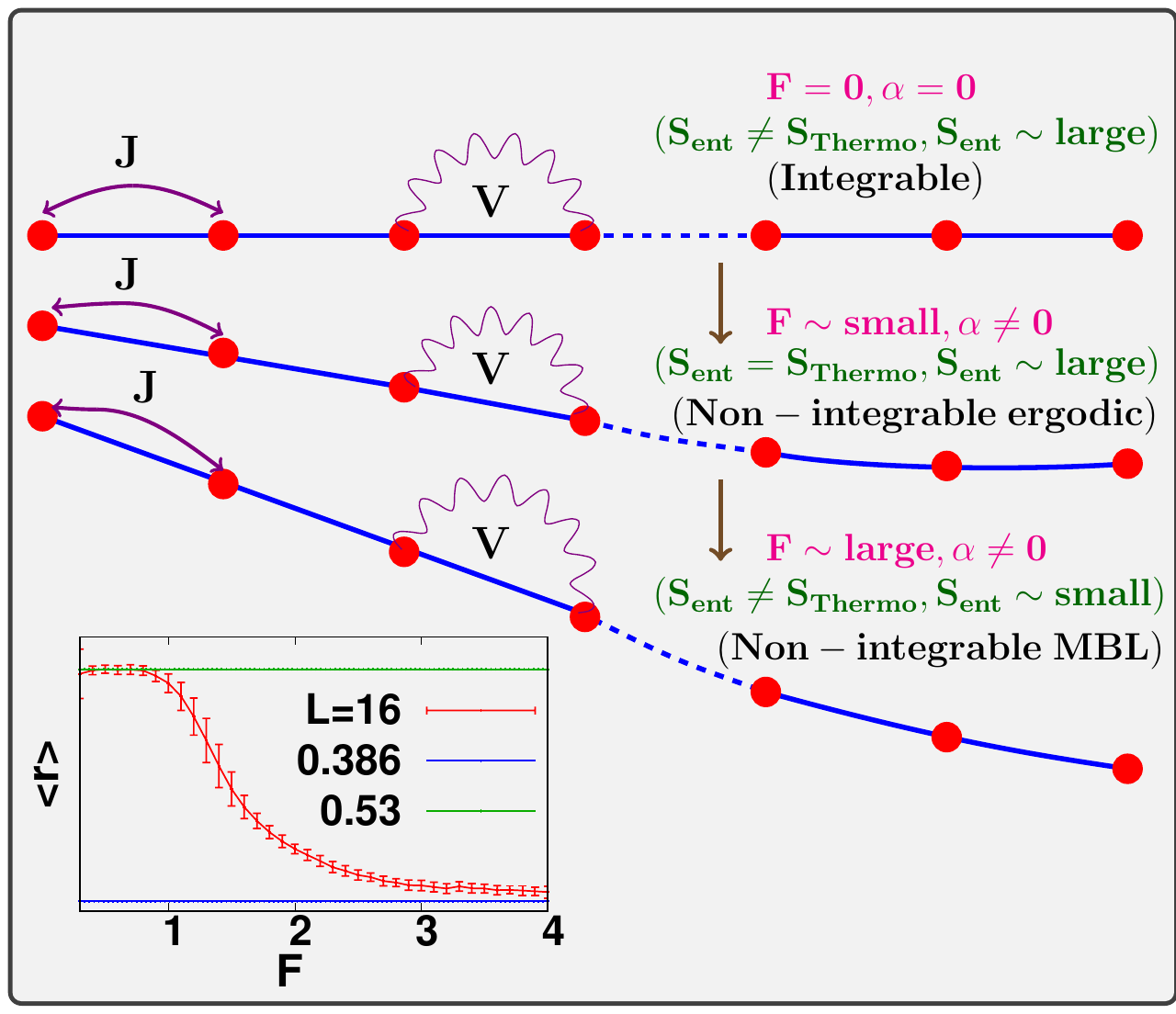}
\caption{A schematic representation of the model with our main
  findings. In the non-integrable ergodic phase, the entanglement
  entropy matches with the thermodynamic entropy, while in the
  integrable and non-integrable MBL phases, it differs from the
  thermodynamic entropy. In the non-integrable MBL phase, the
  magnitude of the entanglement entropy is significantly smaller. The
  arrow represents the direction of increasing field strength. The
  inset shows the mean level-spacing ratio (averaged over different
  values of $\alpha$) as a function of the field strength. The other
  parameters are: $L=16, V=1.0$ and filling factor $=0.5$.}
\label{ravS}
\end{figure}
The phenomenon of many-body localization
(MBL)~\citep{PhysRevLett.95.206603,basko2006metal,PhysRev.109.1492,nandkishore2015many,pal2010many}
in which interactions fail to destroy Anderson localization ( caused
by random disorder) has created considerable excitement. The MBL phase
is believed to exhibit properties similar to those of integrable
systems~\citep{vosk2013many,huse2014phenomenology,vasseur2016nonequilibrium,RevModPhys.91.021001,imbrie2017local,serbyn2013local,chandran2015constructing}. In
particular, although the ETH criterion is known to be satisfied by
generic, non-integrable
systems~\citep{khatami2012quantum,he2013single,tang2015quantum,beugeling2014finite,PhysRevE.81.036206,kim2014testing,PhysRevE.82.031130,PhysRevLett.108.110601,PhysRevE.87.012125},
a violation of the ETH is expected for integrable, and therefore MBL
systems~\citep{PhysRevLett.103.100403,deutsch2013microscopic,mondaini2015many,PhysRevE.85.050102}.
The expectation value of any local observable in these systems
fluctuates wildly for nearby eigenstates. Integrable
  systems are exactly solvable and have an extensive number of local
  conserved currents~\citep{sutherland2004beautiful}, which do not
  evolve in the course of time and hence, prevent the system from
  thermalization. Similarly, MBL systems have conserved quasi-local integrals of
  motion which help to retain the memory of the initial state~\cite{RevModPhys.91.021001,imbrie2017local,serbyn2013local,chandran2015constructing}.

Most MBL systems have in-built disorder
  \citep{luitz2015many,nandkishore2015many,iyer2013many}.  Recent
  work~\citep{schulz2019stark,van2019bloch} has proposed that a stable
  MBL-like phase may be obtained in a clean (disorderless) interacting
  system subjected to an electric field and a confining/disordered
  potential. The additional potential turns out to be essential as in
  its absence, the MBL phase cannot be
  obtained~\cite{schulz2019stark,van2019bloch,moudgalya2019thermalization,taylor2019experimental}. This
many-body system is known to exhibit a rich phase diagram. In the
absence of both electric field and curvature term, this model is
integrable, while a finite value of either of these external
potentials breaks the integrability. Further in the region of broken
integrability it shows a transition from the ergodic to the MBL phase
on varying the strength of the electric field. Thus it provides a good
test bed to characterize various phases: integrable, non-integrable
ergodic and non-integrable MBL phases.  As opposed to a standard
disordered system, a clean system could potentially be realized
experimentally with greater ease, while still using the already
available methods~\citep{schreiber2015observation,choi2016exploring,
  smith2016many,kondov2015disorder,bordia2017probing}.

In this article, we demonstrate the profitability of a study of the
relationship between thermodynamic entropy and entanglement entropy to
characterize various phases.  Although our technique is, in principle,
more general, we concentrate on the concrete case of the above
disorder-free model. We find that for a small subsystem, the
entanglement entropy of each eigenstate matches with the thermodynamic
entropy, provided the system is tuned in the non-integrable ergodic
phase and satisfies the ETH criterion. However in the integrable and
non-integrable MBL phases, the entanglement entropy shows large
fluctuations for nearby eigenstates, and also differs from the
thermodynamic entropy. The difference between the thermodynamic
entropy and the entanglement entropy increases on varying the strength
of the electric field due to the strong localization from the electric
field which leads to a smaller entanglement entropy.  Further tests
are done from an alternative perspective by studying the dynamics of
average particle number in the subsystem. In the long time limit, the
saturation value of the observable in the non-integrable ergodic phase
matches with the results predicted by the diagonal ensemble and the
microcanonical ensemble, while in the non-integrable MBL phase the
saturation value matches with the diagonal ensemble result but differs
from the microcanonical ensemble result.

\section{Model Hamiltonian}
We consider the clean,  spinless fermionic Hamiltonian with $L$ sites ~\cite{schulz2019stark}:
\begin{eqnarray}\label{eq1}
H=-J\sum_{j=0}^{L-2}(c_{j}^{\dagger}c_{j+1}+c_{j+1}^{\dagger}c_{j})-F\sum_{j=0}^{L-1} j (n_{j}-\frac{1}{2})\qquad \nonumber \\ + \alpha \sum_{j=0}^{L-1} \frac{j^2}{(L-1)^2} (n_{j}-\frac{1}{2}) + V\sum_{j=0}^{L-2} (n_j-\frac{1}{2})(n_{j+1}-\frac{1}{2}),\nonumber
\\
\end{eqnarray}
where  $c, c^\dagger$ are the fermionic operators, $F$ is the linear electric field, $\alpha$ is the curvature term
and $V$ is the nearest neighbor interaction. The form of the  curvature term provides a slight non-linearity in the overall onsite potential. The lattice constant is
kept at unity and natural units ($J=\hbar=e=1$) are adopted for all the
calculations.  In the non-interacting limit $(V=0)$ with $\alpha= 0$,
the above Hamiltonian yields the Wannier-Stark ladder
characterized by an equi-spaced energy spectrum proportional to the
electric field strength, and where all the single particle eigenstates
are localized~\citep{krieger1986time,wannier1960wave}. Furthermore,
the dynamics governed by this Hamiltonian gives rise to oscillatory
behavior which is known as Bloch
oscillations~\citep{bouchard1995bloch,mendez1993wannier,hartmann2004dynamics,PhysRevB.98.045408,PhysRevB.99.155149}.
When interactions are included, the model is integrable in the absence
of both the static field and the curvature term ($F=0, \alpha
=0$). The integrability is broken by a non-zero value of either the
field $F$ or the curvature $\alpha$. When the field $F$ is varied while
keeping $\alpha$ fixed at a non-zero value, the system undergoes a
transition from a delocalized (ergodic) phase at small field strengths
to the MBL phase~\citep{schulz2019stark,van2019bloch} at large field
strengths. The inset of Fig.~\ref{ravS} carries a plot of the mean level spacing ratio~\citep{oganesyan2007localization}
(averaged over the curvature parameter $\alpha$) as a function of the field, indicating a change of statistics~\citep{atas2013distribution} from
Wigner-Dyson to Poisson.

\section{ETH and thermodynamic entropy}
For an isolated quantum system described by a Hamiltonian $H$, the time
evolution of any initial state is given by
\begin{equation}\label{eq3}
|\psi(t)\rangle = e^{-iHt}|\psi (0)\rangle = \sum_{n} c_n e^{-i\epsilon_n t} |n\rangle,
\end{equation} 
where $\epsilon_n$ and $|n\rangle$ are the eigenvalues and the
eigenstates of the Hamiltonian respectively. The information of the
initial state is encoded into the coefficients $c_n$. For any operator
$\hat{\mathcal{O}}$ the expectation value after any time $t$ is given
by
\begin{equation}\label{eq4}
\langle\hat{\mathcal{O}}(t)\rangle = \langle \psi(t)|\hat{\mathcal{O}}(t)|\psi(t)\rangle.
\end{equation}
Using Eq.~\ref{eq3}, this simplifies to 
\begin{equation}\label{eq5}
\langle\hat{\mathcal{O}}(t)\rangle = \sum_{n} |c_n|^2 \mathcal{O}_{nn} + \sum_{m\neq n} c^{*}_{m}c_n e^{i(\epsilon_m - \epsilon_n)t} \mathcal{O}_{mn}, 
\end{equation}
where $\mathcal{O}_{mn}$ are the matrix elements of the operator
$\hat{\mathcal{O}}$ in the eigenbasis of the Hamiltonian $H$. It can
be seen from Eq.~\ref{eq5} that in the long time limit ($t\to\infty$),
generically (in the absence of degeneracy) the second term goes to zero and the expectation value of the
observable saturates to the value predicted by the diagonal ensemble:
\begin{equation}\label{eq5a}
\langle\hat{\mathcal{O}}(t\to \infty)\rangle = \langle\hat{\mathcal{O}}_{DE}\rangle = \sum_{n} |c_n|^2 \mathcal{O}_{nn}. 
\end{equation}
Hence the system retains the memory of the initial state through the coefficients
$c_n$, and does not follow the ergodic hypothesis.
\begin{figure}[t]
\includegraphics[scale=0.56]{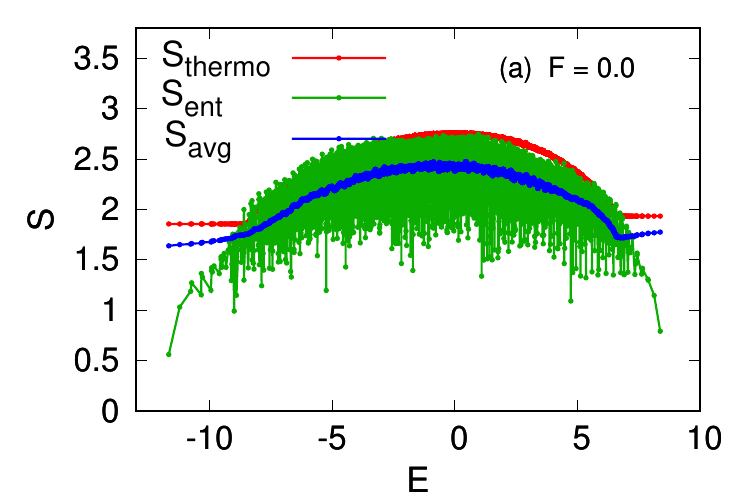}
\includegraphics[scale=0.56]{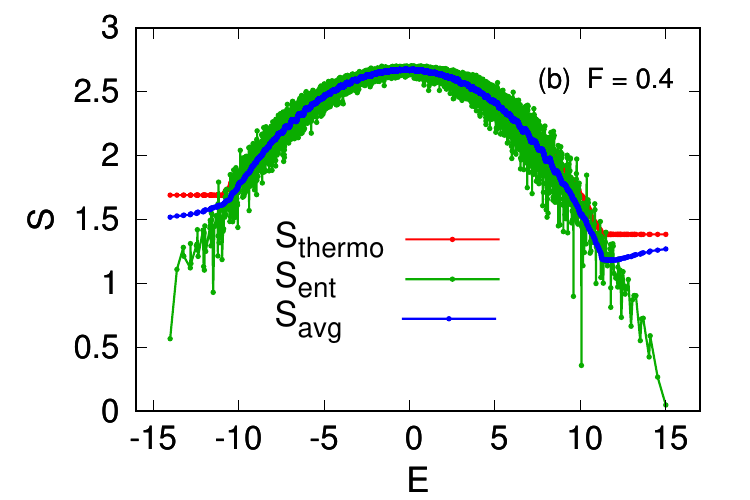}\\
\includegraphics[scale=0.56]{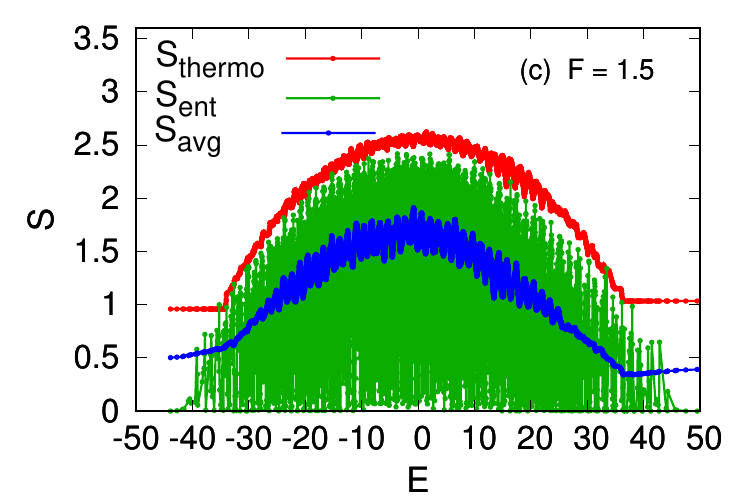}
\includegraphics[scale=0.56]{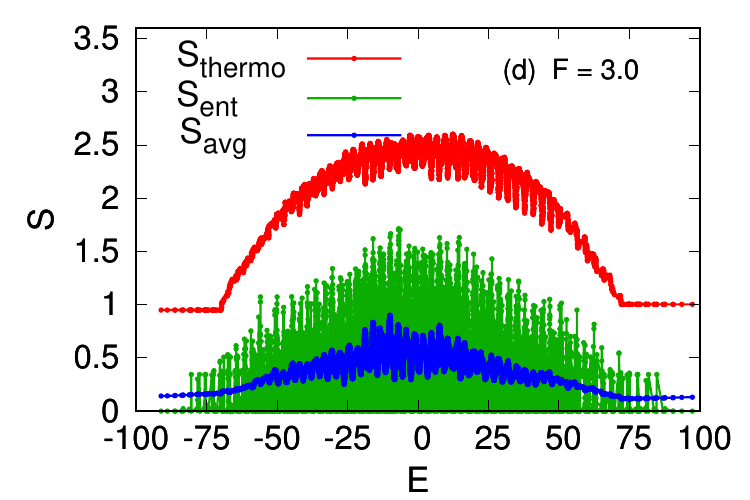}
\caption{The entanglement entropy of each energy eigenstate and the
  corresponding thermodynamic entropy. (a) Integrable phase ($F=0$,
  $\alpha =0$): the entropy of nearby eigenstates fluctuates wildly with
  a finite difference between average entropy and the microcanonical
  average. (b,c,d) Non-integrable phase : with $\alpha=1.0$ and
  $F=0.4,1.5,\ \text{and}\ 3.0$ respectively. We obtain agreement between
  the entanglement entropy with its corresponding thermodynamic
  entropy in the ergodic phase $(F=0.4)$ satisfying ETH
  while the ETH is violated on increasing the value of field strength
  (going into the MBL phase). The other
  parameters are: $L=16, V=1.0$ filling factor $=0.5$, and subsystem size $m = 4$.}
\label{enteig}
\end{figure} 

Thermalization in isolated quantum many body systems happens via the
mechanism of ETH, which implicitly involves the assumption that the
diagonal elements of the operator $\hat{\mathcal{O}}$ change slowly
with the eigenstates. Specifically, the off-diagonal elements
$\mathcal{O}_{mn}$, and the difference in the neighboring diagonal
elements: $\mathcal{O}_{n+1,n+1}-\mathcal{O}_{n,n}$ are exponentially
small in $\mathcal{N}$, with $\mathcal{N}$ being the system size.
With this assumption, the diagonal ensemble result (Eq.~\ref{eq5a})
saturates to a constant value as the matrix elements
$\mathcal{O}_{nn}$ are effectively constant over a given energy
window.

Now considering the micro-canonical ensemble, the average value of the
same observable can be written as
\begin{equation}\label{eq5b}
\langle\hat{\mathcal{O}}_{ME}\rangle = \frac{1}{N_{\text{states}}}\sum_{n=1}^{N_{\text{states}}} \mathcal{O}_{nn},
\end{equation}
where $N_{\text{states}}$ is the number of states in a given energy  shell. Imposing the assumption of ETH, this also saturates to a constant value. Thus in the long time limit, the system thermalizes and the observable saturates to a thermal value predicted by the micro-canonical ensemble~\citep{rigol2008thermalization,kim2014testing,PhysRevLett.108.110601}.

Under these conditions the expectation value of the operator
$\hat{\mathcal{O}}$ in the energy eigenstate characterized by the
density matrix $\rho_E \equiv |E\rangle\langle E|$ is the same as the micro-canonical average of the
same operator:
\begin{equation}\label{eq6}
\text{Tr}(\rho_E \hat{\mathcal{O}})=\text{Tr}(\rho_{\text{micro},E}\hat{\mathcal{O}}),
\end{equation}
where the microcanonical density matrix is defined as 
\begin{equation}\label{eq7}
\rho_{\text{micro},E_0} = \frac{1}{N_{\text{states}}}\sum_{E_0<E< E_0+\Delta E} |E\rangle\langle E|,
\end{equation}
where $N_{\text{states}}$ is the number of states available in the energy window $\Delta E$.  
\begin{figure}[t]
\includegraphics[scale=1.1]{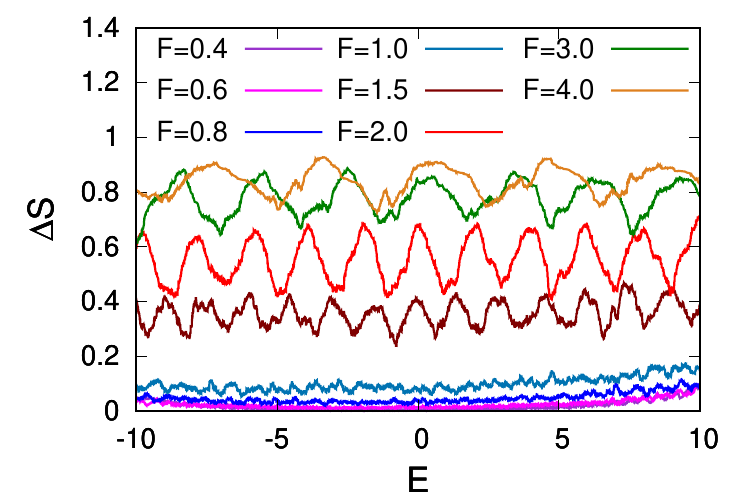}
\caption{The difference between the thermodynamic entropy and the average entropy as a function of energy. Only the central part of the spectrum ($E\in[-10:10]$) is shown for various values of the field strength. In the ergodic phase the difference is almost zero while in the MBL phase the difference is much larger. The other
  parameters are: $L=16, \alpha = 1.0, V=1.0$ filling factor $=0.5$, and subsystem size $m = 4$.}
\label{avgentdiff}
\end{figure}
\begin{figure*}[t]
\includegraphics[scale=0.77]{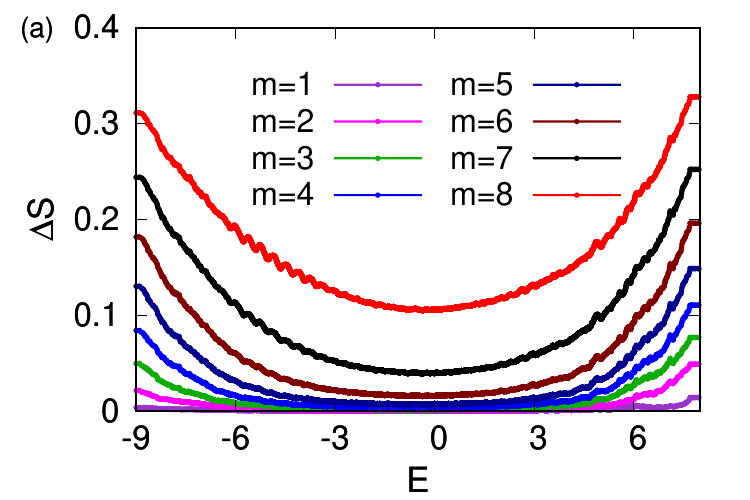}
\includegraphics[scale=0.77]{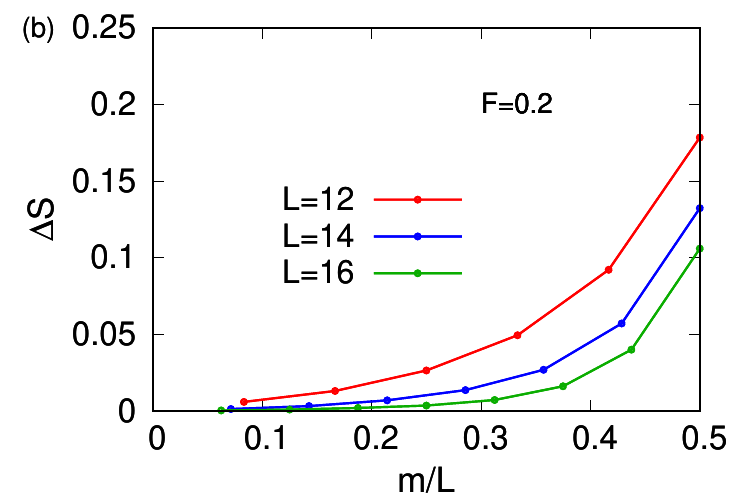}
\includegraphics[scale=0.77]{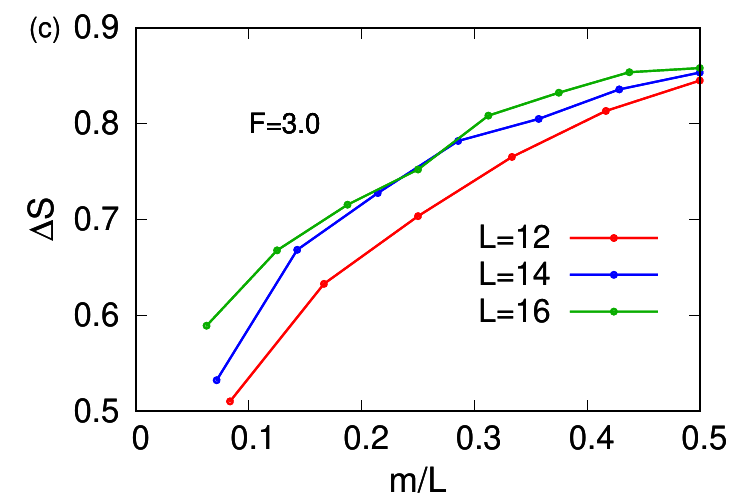}
\caption{(a) Difference between the thermodynamic entropy and the
  average entropy (average carried out over $100$ nearest eigenstates
  in both cases) as a function of energy for different subsystem sizes
  in the ergodic phase($F=0.2$). A better thermalization can be seen
  for smaller subsystem sizes. (b,c) The finite size scaling of the
  difference of thermodynamic and average entropy (for a single
  eigenstate located at the middle of spectrum) as a function of the
  subsystem size in both ergodic and MBL phases. The other
  parameters are: $L=16, \alpha = 1.0, V=1.0$ filling factor $=0.5$. }
\label{diff_m} 
\end{figure*}
For a composite system $(A+B)$ characterized by the density matrix
$\rho$, the entanglement entropy of a subsystem $A$ is defined as:
$S_{\text{Ent}}=-\text{Tr}(\rho_{A}\text{ln}\rho_{A})$, where $\rho_A
= \text{Tr}_B \rho$, is the reduced density matrix of the subsystem
$A$ taken after tracing out the degrees of freedom of the other
subsystem $B$. On the other hand, the thermodynamic entropy from a
microcanonical ensemble is defined as:
$S_{\text{thermo}}=-\text{Tr}(\rho_{\text{micro}}\text{ln}\rho_{\text{micro}})$.
The criterion of ETH is
extended~\citep{deutsch2013microscopic,deutsch2010thermodynamic} by
asking whether the entanglement entropy of a small subsystem taken out
of a large system in an eigenstate with energy $E_0$ is equal to the
thermodynamic entropy computed from the micro-canonical density matrix
(Eq.~\ref{eq7}) with the same energy $E_0$. Positing an ETH-like
equation where $\rho_{\text{micro}}$ is replaced by $\rho_A$ we ask if
the condition
\begin{equation}\label{eq11}
S_{\text{thermo}} = -\text{Tr}(\rho_{A}\text{ln}\rho_{A})=-\text{Tr}(\rho_{A,\text{micro}}\text{ln}\rho_{A,\text{micro}})
\end{equation}
holds. Here, $\rho_{A,\text{micro}}$ is the reduced density matrix corresponding to the density matrix $\rho_{\text{micro}}$. For the subsystem $A$, this can be calculated by tracing out the degree of freedom of the remaining part: $\rho_{A,\text{micro}}= \text{Tr}_{B}(\rho_{A,\text{micro}})$.  

Although the above criterion is analogous to the standard ETH
one (Eq.~\ref{eq6}) the logarithmic factor $\text{ln}\rho_A$ is not an
observable quantity, thus making it an independent characteristic of
thermalization.

\section{Results and discussion}
\subsection{Statics} The model considered contains three regimes of
interest: the integrable phase, the non-integrable ergodic phase and
the non-integrable MBL phase. We employ numerical exact
diagonalization of the model (Eq.~\ref{eq1}) for a system size upto
$L=16$ with the filling factor set to half filling.  We also define the subsystem $A$ as consisting of first $m$ sites out of the $L$ sites. We test the equivalence of
the thermodynamic entropy and entanglement entropy (Eq.~\ref{eq11}) in these distinct
phases. We compute the entanglement entropy for a small
subsystem  ($m=4$) for all the eigenstates and plot it in
Fig.~\ref{enteig}. The thermodynamic entropy for
all the eigenstates is also plotted by considering the microcanonical
density matrix (Eq.~\ref{eq7}), followed by tracing out the degrees of
freedom of the complement of the subsystem. Since the energy spectrum fans out as
a function of the electric field strength, we average the density
matrix over $N_\text{states}=100$ nearest-neighbor eigenstates to compute the
thermodynamic entropy. Furthermore, the average
entanglement entropy $S_\text{avg}$ (average of the entanglement entropy of $100$
nearby eigenstates) is also plotted in the same figure.

In the integrable case ($F,\alpha = 0$), the thermodynamic entropy
differs from the entanglement entropy with the latter having a lot of
fluctuations. However, for the parameters in the ergodic phase, nice
agreement is found between the thermodynamic entropy and entanglement
entropy, which signifies the validity of ETH in this phase. When the
system is tuned on the border ($F=1.5$), the entanglement entropy also
shows fluctuations due to a mixture of both volume law and area law
scaling states. This in-between phase has been called the ``S-phase"
~\citep{xu2019butterfly}. For the parameters in the MBL region, the
entanglement entropy shows wild fluctuations and the thermodynamic
entropy is also different from the entanglement entropy, which suggests
the breakdown of ETH in the MBL phase.  It is interesting to note that
even though both integrable and non-integrable MBL phases violate the
ETH, the \emph{magnitude} of entanglement is considerably lower in the latter, due to the underlying localization.

It is useful to consider the difference between thermodynamic entropy
and the average entanglement entropy:
\begin{equation}
\Delta S =
\frac{S_{\text{thermo}}-S_{\text{avg}}}{S_{\text{thermo}}}.
\end{equation}
The difference between the
thermodynamic entropy and the entanglement entropy ($\Delta S$)
increases on increasing the electric field strength. The entropy for a
part of the spectrum ($E\in[-10:10]$) is plotted in
Fig.~\ref{avgentdiff} for various values of the field strengths. In
the ergodic phase the difference is close to zero signifying the
validity of ETH while a finite difference in the MBL phase shows the
violation of ETH.
\begin{figure}[t]
\includegraphics[scale=1.1]{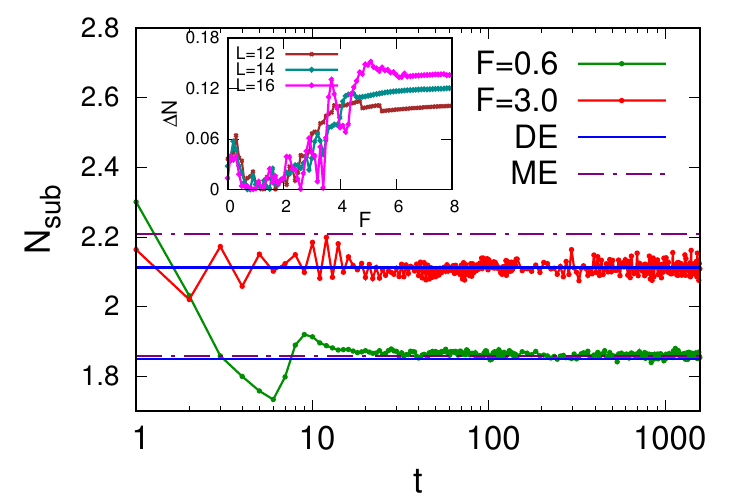}
\caption{Quench dynamics: In the non-integrable ergodic phase
  ($F=0.6$), the long time saturation value of the average number of
  particles in the subsystem matches with those of the diagonal ensemble
  and the micro-canonical ensemble. In the non-integrable MBL
  phase ($F=3.0$) on the other hand, the saturation value matches with the result of the
  diagonal ensemble while it differs from that of the
  micro-canonical ensemble. The inset shows the normalized difference
  between the diagonal ensemble result and the micro-canonical
  ensemble result as a function of field strength for the same initial
  state. The value is close to zero in the non-integrable ergodic
  phase while a finite difference is obtained in the non-integrable
  MBL phase. The other parameters are: $L=16, \alpha = 1.0, V=1.0$
  filling factor $=0.5$, and subsystem size $m = 4$.}
\label{quench}
\end{figure} 

Finally, we test the equivalence of thermodynamic entropy
  and entanglement entropy on varying the subsystem size. For each
  eigenstate, Fig.~\ref{diff_m} shows the difference between these two
  for various values of subsystem size. It can be seen that for
  smaller subsystems the difference tends to zero, hence the
  smaller the subsystem the better is the thermalization~\cite{PhysRevLett.119.020601,PhysRevLett.121.220602}. The other
  two figures in Fig.~\ref{diff_m} show the finite size scaling of
  this difference but for a single eigenstate located at the center of
  the spectrum. It can be seen that for a smaller fraction $m/L$ the
  difference goes to zero and thus shows the validity of ETH for these
  fractions. On the other hand, in the MBL phase, this difference is
  found to increase on increasing the system size as well as the
  subsystem sizes.

\subsection{Quench dynamics} A complementary understanding of the
distinction between the various phases is afforded by a study of the
long time behavior of the system under time evolution. As evident from
Eq.~\ref{eq5}, the dynamics of any observable has two parts: the first
part is the same as the result predicted by the diagonal ensemble while
the second part gives the fluctuations around it. In the long time limit, 
the observable, in general, equilibrates to the diagonal ensemble value. However this does not imply the
thermalization of the observable. An observable is said to thermalize
if the result of the diagonal ensemble matches with the result
predicted by any thermal ensemble such as micro-canonical or canonical.

We consider the average number of particles in the
subsystem~\citep{PhysRevLett.115.186601}: $\hat{\mathcal{O}} =
\sum_{i=1}^{m} \hat{N}_i$, where $\hat{N}_{i}=c_{i}^{\dagger}c_i$ is
the number operator at site $i$.  The initial state is taken as a
charge density wave state (where all the even sites are occupied and
odd sites are empty), and the dynamics is governed by the final
Hamiltonian (Eq.~\ref{eq1}). The prescription for obtaining the micro-canonical density
matrix is as follows. We first calculate the average
energy of the initial state: $E_{\text{ini}} = \langle
\psi_0|H|\psi_0\rangle$.  Next we obtain the eigenstate closest to this energy.
By taking $100$ nearest neighbor eigenstates around the obtained
state, we then construct the micro-canonical density matrix.
\begin{figure}[t]
	\includegraphics[scale=1.2]{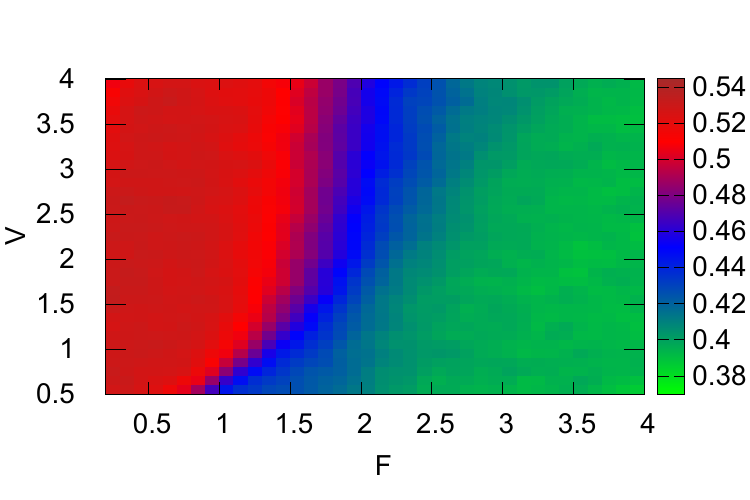}
	\caption{Surface plot of the level statistics as a function of both field strength $(F)$ and the interaction strength $(V)$.  The other parameters are: $L=16, \alpha=1.0$. }
	\label{rav}
\end{figure}

We present data for the dynamics of the above observable in
Fig.~\ref{quench}, comparing against the values predicted by the
diagonal and micro-canonical ensembles.  In the ergodic phase, the
long time limit of the expectation value of the observable is in
agreement with that predicted by both the diagonal ensemble and the
micro-canonical ensemble, which in turn implies thermalization and the
validity of ETH in this phase. On the other hand, in the MBL phase the
saturation value is the same as predicted by the diagonal ensemble but
it differs from the micro-canonical ensemble result suggesting the
lack of thermalization in the MBL phase. To study the difference
between the diagonal and micro-canonical ensemble results, we define
the following normalized difference:
\begin{equation}\label{eq14}
\Delta N = \frac{|N_{\text{DE}} - N_{\text{ME}}|}{|N_{\text{ME}}|},
\end{equation}
where $N_{\text{DE}}$ and $N_{\text{ME}}$ are the expectation values
of the observable $\hat{\mathcal{O}}$, calculated from the diagonal
ensemble and micro-canonical ensemble respectively.  The inset shows
the normalized difference $\Delta N$ (Eq.~\ref{eq14}) as a function of
electric field strength for the same initial state. The value is close
to zero in the non-integrable ergodic phase while a finite difference
is obtained in the non-integrable MBL phase.

\subsection{Variation of interaction strength}
The nature of the phase obtained also depends on the interaction
strength. Fig.~\ref{rav} shows the surface plot of the average level
spacing as a function of both field strength and interaction
strength for a fixed value of the curvature term ($\alpha = 1.0$). It
can be seen that on increasing the interaction strength, the ergodic
region extends, thus we expect the equivalence of the entanglement
entropy and the thermodynamic entropy to hold in this extended region.

\section{Summary and Conclusions}  To summarize, we test the validity
of ETH in an interacting system subjected to a static electric
field. For small electric field strength this model shows ergodic
behavior while for sufficiently strong electric field it exhibits
MBL. In the limit of zero electric field and curvature strength, the
model is integrable. We find that in the ergodic phase, the
entanglement entropy of the states following a volume law of scaling
matches with the corresponding thermodynamic entropy thus satisfying
the ETH criterion, while in the MBL phase, the entanglement entropy
fluctuates wildly from eigenstate to eigenstate, and also differs from the thermodynamic
entropy. Since the MBL phase possesses low entanglement, a clear
distinction is obtained between the integrable and the MBL phase from
the point of view of the ETH. As reported
earlier~\citep{deutsch2013microscopic}, a striking distinction between
integrable and non-integrable systems is the presence of large
eigenstate-to-eigenstate fluctuations in the expectation value of any
observable in the integrable case. In support of the argument that the MBL phase is similar
to integrable systems, we find that indeed, the MBL phase is also characterized by large
flucutations in entanglement entropy across adjacent eigenstates. However, in contrast to
the integrable phase, the \emph{magnitude} of entanglement is significantly lower in the MBL
phase. Moreover, the difference between the average entropy and the thermodynamic entropy
increases on going deep into the localized phase.

We further verify the above arguments from a dynamical perspective by
studying the dynamics of average number of particles in the subsystem
starting from a charge density wave type of initial state. We find
that in the ergodic phase the saturation value obtained from the dynamics, the result predicted
by the diagonal ensemble as well as the micro-canonical ensemble result
match with each other, implying that the system thermalizes in the
long time limit. In the MBL phase on the other hand, the saturation value matches with
the result predicted by the diagonal ensemble, but differs from
that predicted by the micro-canonical ensemble. This signifies
the lack of thermalization or ETH in the MBL phase.

\section*{Acknowledgment}
We are grateful to the High Performance Computing (HPC) facility at
IISER Bhopal, where large-scale calculations in this project were
run. A.S is grateful to SERB for the grant (File Number:
CRG/2019/003447), and for financial support via the DST-INSPIRE
Faculty Award [DST/INSPIRE/04/2014/002461]. D.S.B acknowledges PhD
fellowship support from UGC India. We are grateful to Josh Deutsch and
Sebastian W\"{u}ster for their comments on the manuscript. We thank
Ritu Nehra for help with the schematic diagram.

\bibliography{ref}
\newpage
\onecolumngrid
\appendix
\section{Scaling of Entanglement Entropy and Thermodynamic Entropy }
\begin{figure}[!h]
	\includegraphics[scale=1.0]{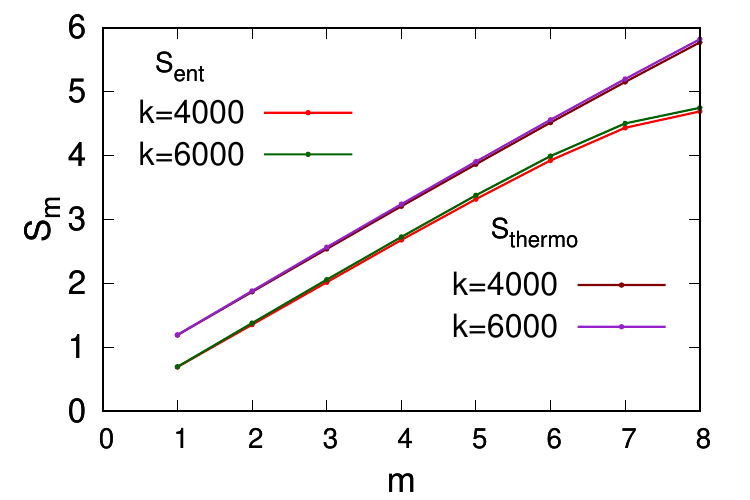}
	\includegraphics[scale=1.0]{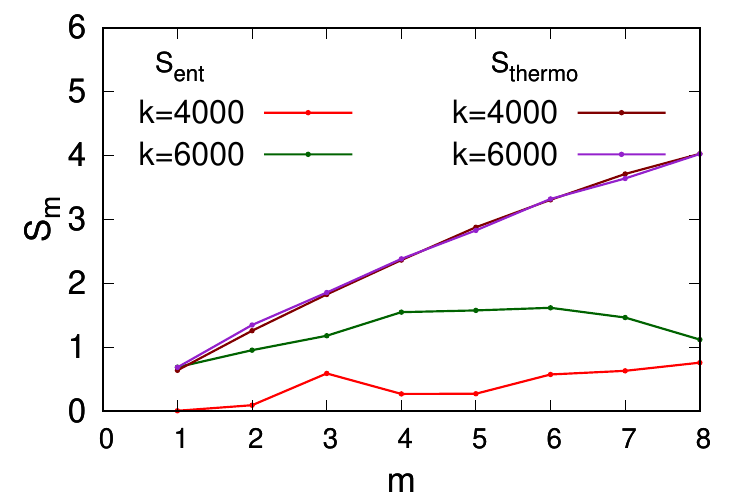}
	\caption{The scaling of entanglement entropy and thermodynamic entropy as a function of subsystem size for two different eigenvectors in the middle of the spectrum. In the ergodic phase ($F=0.2$), both follows a volume law scaling (Left). The thermodynamic entropy is shifted by an amount $0.5$ from the clarity purpose. In the MBL phase $(F=3.0)$, only the thermodynamic entropy shows a volume law scaling (right). The other parameters are: $L=16, \alpha=1.0\  \text{and}\ \  V=1.0$. }
	\label{scale_ent}
\end{figure}

In this appendix, we provide the scaling of the entanglement entropy and the thermodynamic entropy of two random states from the middle of the spectrum as a function of subsystem size. In the ergodic phase ($F=0.2$), both the entropy matches with each other and follows a volume law scaling, while in the MBL phase $(F=3.0)$, only the thermodynamic entropy shows a volume law scaling (Fig.~\ref{scale_ent}).

\end{document}